# Abnormal behavior of preferred formation of cationic vacancy from the interior in γ-GeSe monolayer with the stereo-chemical antibonding lone-pair state

Changmeng Huan,[a,b] Yongqing Cai,*[c] Devesh R. Kripalani,[d] Kun Zhou [d] and Qingqing Ke *[a,b]

[a] School of Microelectronics Science and Technology, Sun Yat-sen university, Zhuhai 519082, China

[b] Guangdong Provincial Key Laboratory of Optoelectronic Information Processing Chips and Systems, Sun Yat-sen University, Zhuhai 519082, China

[c] Joint Key Laboratory of the Ministry of Education, Institute of Applied Physics and Materials Engineering, University of Macau, Taipa, Macau, China

[d] School of Mechanical and Aerospace Engineering, Nanyang Technological University, 50 Nanyang Avenue, Singapore 639798, Singapore

* Corresponding authors

E-mail: yongqingcai@um.edu.mo; keqingq@mail.sysu.edu.cn

# Abstract

Two-dimensional (2D) materials tend to have the preferably formation of vacancies at the outer surface. Here, contrary to the normal notion, we reveal a type of vacancy that thermodynamically initiates from the interior part of the 2D backbone of germanium selenide (γ-GeSe). Interestingly, the Ge-vacancy ($V_{Ge}$) in the interior part of γ-GeSe possesses the lowest formation energy amongst the various types of defects considered. We also find a low diffusion barrier (1.04 eV) of $V_{Ge}$ which is a half of those of sulfur vacancy in $MoS_2$. The facile formation of mobile $V_{Ge}$ is rooted in the antibonding coupling of the lone-pair Ge 4s and Se 4p states near the valence band maximum, which also exists in other gamma-phase MX (M=Sn, Ge; X=S, Te). The $V_{Ge}$ is accompanied by a shallow acceptor level in the band gap and induces strong infrared light absorption and p-type conductivity. The $V_{Ge}$ located in the middle cationic Ge sublattice is well protected by the surface Se layers – a feature that is absent in other atomically thin materials. Our work suggests that the unique well-buried inner $V_{Ge}$, with the potential of forming structurally protected ultrathin conducting filaments, may render the GeSe layer an ideal platform for quantum emitting, memristive, and neuromorphic applications.

# Introduction

Group IV monochalcogenides MX (M = Ge, Sn, Pb; X = S, Se, Te) are a family of emerging 2D layered materials which are gaining extensive interests in thermoelectric,[1,] [2] photocatalysis,[3, 4] and ferroelectric[5, 6] fields. Among the various MXs, GeSe deserves particular attention, owing to recent papers reporting on its outstanding photovoltaic efficiency,[7] excellent optoelectronic properties,[8] and electric-field induced room temperature antiferroelectric-ferroelectric phase transition,[9] as well as its inherent chemical stability and environmental sustainability.

Thus far, most of the studies on MX are all based on the well-studied orthorhombic phase (α-phase) with a black-phosphorus-like structure. Recently, a hexagonal phase of GeSe (γ-phase) with two merged blue-phosphorus-like layers has been synthesized on a h-BN substrate by chemical vapor deposition (CVD), showing an intriguing electronic conductivity that is even higher than that of graphite.[10] Unlike theoretical predictions, the deposited γ-GeSe shows an A-B' stacked non-centrosymmetric atomic structure.[11] This new structure has aroused great interests with respect to the strain-tunable electronic structure and spontaneous polarization,[12, 13] van der Waals heterostructures,[14] and fast Li-ion intercalation properties[15] through first-principles calculations.

Point defects are inevitably introduced during the preparation process and have significant effects on the electronic, optical, and magnetic properties of 2D materials.[16-19] For instance, oxygen vacancies in ultrathin $WO_3$ layers can introduce an intermediate band and induce strong infrared light absorption, enabling IR-driven $CO_2$ splitting into CO and $O_2$.[18] Cr or I atom vacancies can induce a ferromagnetic to antiferromagnetic transition in $CrI_3$.[19] Therefore, it is of crucial importance to understand the nature of intrinsic defects and their impact on the properties of γ-GeSe.

In this work, using first-principles calculations, we discovered an abnormal type of defect that is likely to exist in this new phase of GeSe. Through investigating the intrinsic point defects in the hexagonal phase γ-GeSe monolayer, including vacancies, anti-site defects, and interstitials, an energetically favorable formation of the Ge vacancy located at the interior region is predicted, which is distinctive from the overwhelming dominance of anionic vacancies at the surface in existing 2D compounds. The advantage of the spontaneous formation of such inner vacancy voids is apparent as it allows the manipulation of robust and structurally intact ordering of the vacancy for quantum emission or resistive switching. Thus far, achieving similar functionality in other 2D materials has to resort to generating surface defects such as halogen vacancies in transition metal dichalcogenides which tend to be subjected to structural poisoning and environmental disturbance.[20] Through further analysis of its orbital hybridization diagram, we attribute the formation of this facile defect to the significant population of lone-pair Ge 4s and Se 4p states near the valence band maximum, which also exists in other gamma-phase MX (M=Sn, Ge; X=S, Te).

## Results and discussion

**Structures and stability of perfect γ-GeSe**

Monolayer γ-GeSe is a four-atom-thick (Se-Ge-Ge-Se sequence, Fig.1a) monolayer with a buckled honeycomb lattice structure that can be exfoliated from its bulk phase (space group *P6$_3$mc*, Fig.1b).[10] The lattice dynamic stability of monolayer γ-GeSe is confirmed by the lack of imaginary phonon modes in the phonon dispersion spectra, as shown in Fig.1c. Furthermore, its thermodynamic stability is also validated through AIMD calculation (Fig.1d), where the total energy undergoes a small fluctuation, and the structural snapshot at 10 ps shows that the atoms are only slightly displaced near their equilibrium positions with no structural corruption, thus confirming the excellent thermostability of this phase.

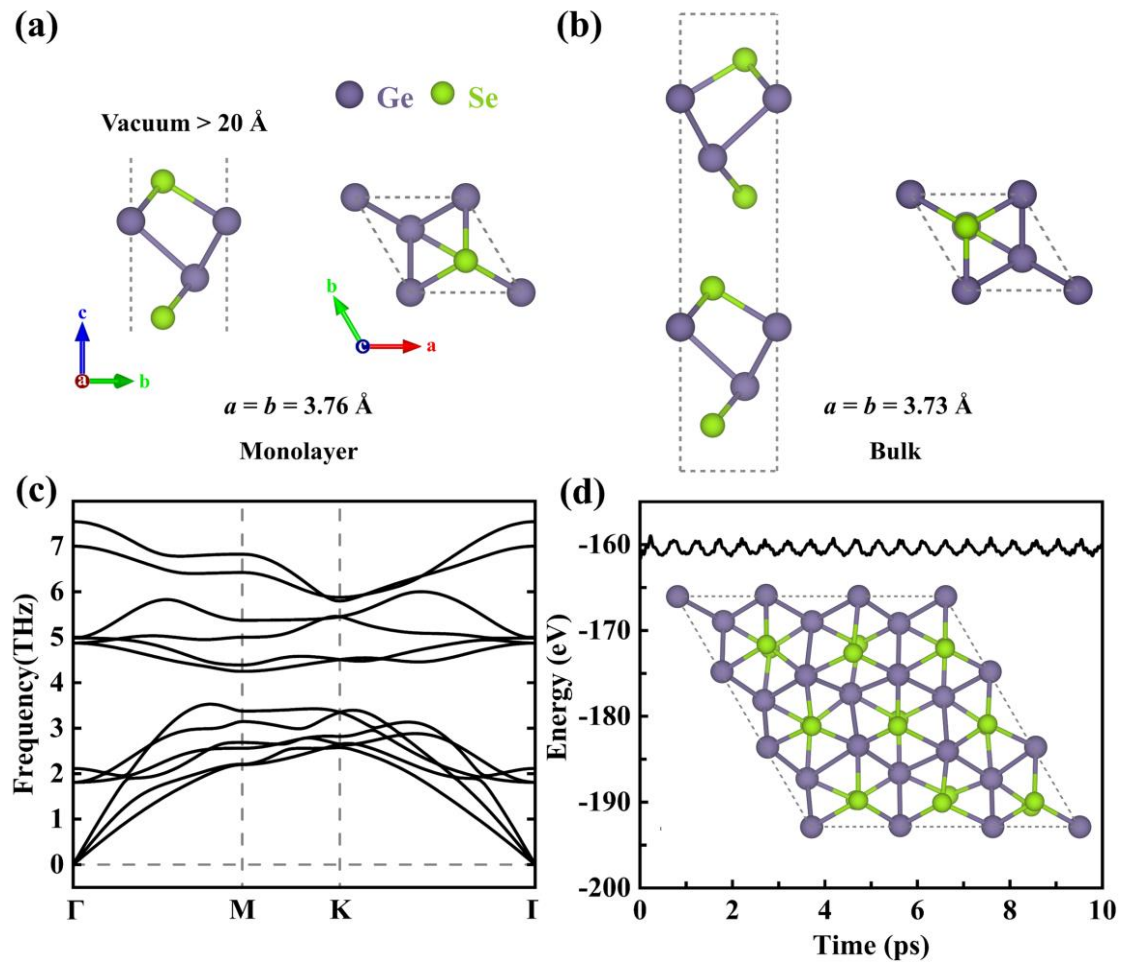

**Fig. 1** Front and top views of the unit cell of (a) monolayer and (b) bulk γ-GeSe, (c) phonon spectra of monolayer γ-GeSe, and (d) evolution of the total energy from AIMD calculation at 300 K and a snapshot of the monolayer supercell at 10 ps.

**Formation of defects in monolayer γ-GeSe**

The point defects in monolayer γ-GeSe were cataloged into three types: (I) vacancy: a single Ge atom vacancy ($V_{Ge}$), a single Se atom vacancy ($V_{Se}$), a double Ge atom vacancy ($V_{2Ge}$), a double Se atom vacancy ($V_{2Se}$), and a Ge-Se vacancy ($V_{Ge-Se}$); (II) anti-site defect: where a Ge atom substitutes an Se atom ($Ge_{Se}$) or an Se atom substitutes a Ge atom ($Se_{Ge}$); (III) interstitial: an interstitial Ge atom ($I_{Ge}$) and an interstitial Se atom ($I_{Se}$). The double vacancies ($V_{2Se}$, $V_{Ge-Se}$, $V_{2Ge}$) are generated by removing the nearest-neighboring atoms without considering any asymmetry in this work. The optimized geometries of defective γ-GeSe monolayers in the neutral charge state are presented in Fig. S1. It is found that all defective supercells show slight structural changes relative to the perfect one, while single-point defects exhibit relatively localized distortions relative to double-point defects.

The structural stability and thermodynamics of the different point defects can be evaluated from their formation energy ($E_f$), which can be defined as[21]

$$E_f = E_d - E_p - \sum n_i \mu_i \qquad (1)$$

where $E_d$ and $E_p$ represent the total energies of the defective and perfect γ-GeSe, respectively. $n_i$ is the number of $i$-type atoms (Ge or Se) added to ($n_i < 0$) or removed from ($n_i > 0$) the supercell, and $\mu_i$ indicates the chemical potential of atoms of type $i$. The range of $\mu_{Ge}$ in GeSe is limited by[22]

$$\mu_{GeSe} - \mu_{Se}^O < \mu_{Ge} < \mu_{Ge}^O \qquad (2)$$

where $\mu_{Se}^O$ and $\mu_{Ge}^O$ represent the Se chemical potential in the Se$_6$ molecular crystal and the Ge chemical potential in pure Ge bulk, respectively. $\mu_{GeSe}$ indicates the total energy per formula unit of monolayer GeSe.

The $E_f$ of all the investigated defects in monolayer γ-GeSe as a function of $\mu_{Ge}$ are plotted in Fig. 2, and the $E_f$ of such defects in monolayer α-GeSe (Fig. S2) are also calculated for comparison. Surprisingly, the V$_{Ge}$ defect shows the lowest $E_f$ over the whole range of Ge chemical potential values despite its formation requiring the breaking of six covalent bonds and the removal of cationic Ge from the inner layers of γ-GeSe. This preferred formation of cationic vacancies is quite different from the case of MoX$_2$ (X = S, Se, Te), where X vacancies in the outer layer are easier to form.[23-25] Furthermore, the formation energy of V$_{Ge}$ (0.92/0.46 eV) in γ-GeSe is even lower than that in the common α-GeSe phase (1.16/0.84 eV), where Ge atoms are located in the outer layer and lie exposed on the surface. The effect of concentration on energetics of vacancy is considered by comparing the $E_f$ of the single vacancy in 2×2, 3×3, 4×4 supercells, and the $E_f$ of the single vacancy and double vacancies in a 3×3 supercell, respectively. The results in Fig. S3 indicate that this abnormal behavior of preferred formation of cationic vacancy from the interior exists in all conditions considered, and the lower the vacancy concentration, the lower the $E_f$ of V$_{Ge}$. In addition, as shown in Fig. S4, the interaction energies ($E_{int}$)[26] of di-vacancies is smaller than that of two separated mono-vacancies, indicating that the separated mono-vacancies in monolayer γ-GeSe tend to combine into vacancy pairs. This is consistent with the scenario in graphene but in contrast to that in MoS$_2$, where separated mono-vacancies are more energetically favored.[25, 27] On the other hand, AIMD simulation has validated the thermodynamic stability of the γ-GeSe supercell containing different defects (Fig. S5). Therefore, Ge vacancies and its clusters are the most likely defect in monolayer γ-GeSe. Certainly, other defects can also be achieved by irradiation and ion implantation, which are widely accepted approaches for the introduction and control of defects.

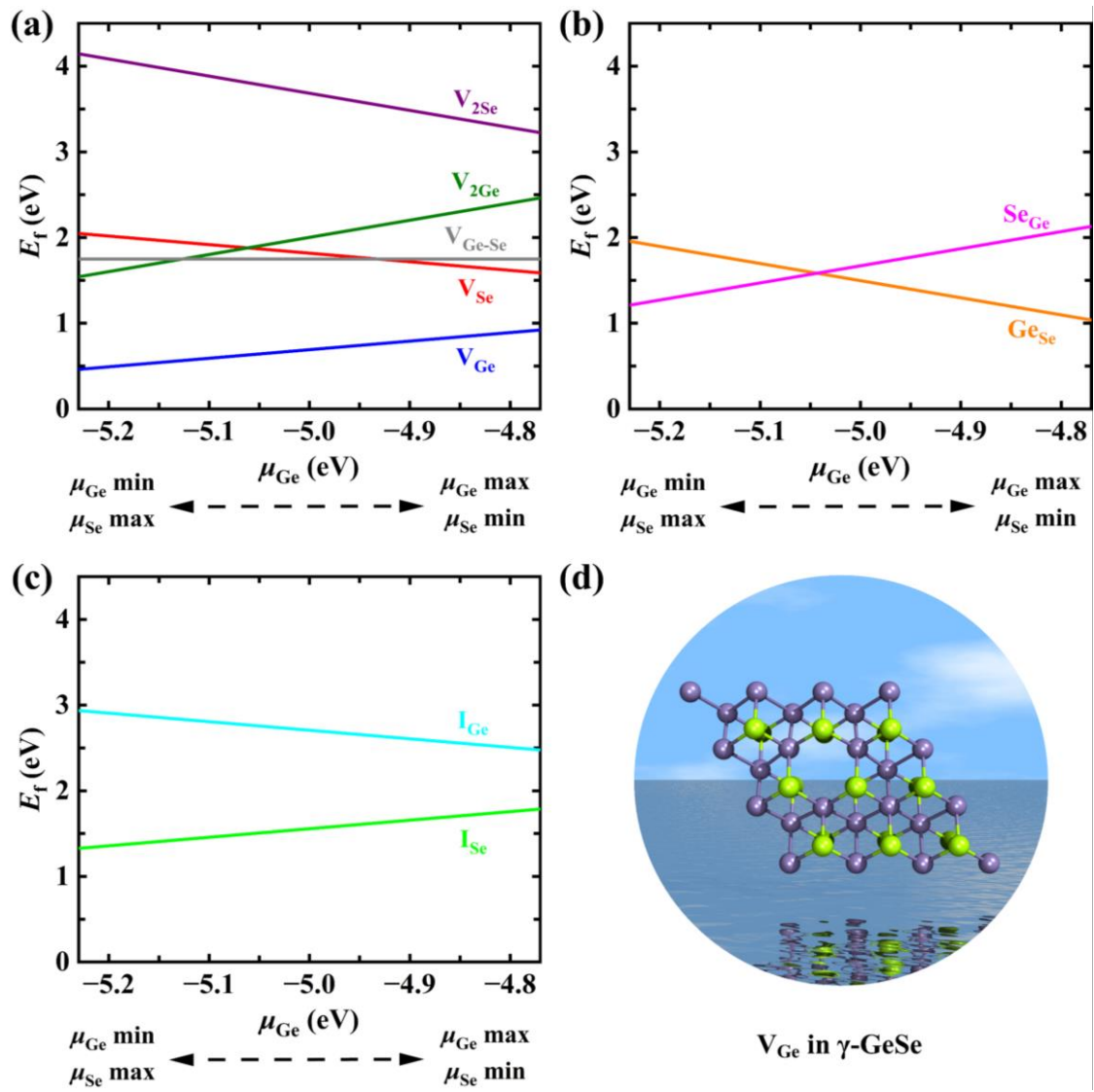

**Fig. 2** Formation energies of monolayer γ-GeSe containing (a) vacancy, (b) anti-site,

and (c) interstitial defects as functions of $\mu_{Ge}$ in the range from −5.23 eV (Ge-poor) to −4.77 eV (Ge-rich). (d) Top view of the most stable defect ($V_{Ge}$) in the γ-GeSe monolayer. Where γ-GeSe can remain stable with respect to the formation of bulk Ge ($\mu_{Ge}$ = −4.77 eV) or the Se$_6$ molecular crystal ($\mu_{Ge}$ = −5.23 eV).

**Mechanism of thermodynamically preferred Ge vacancy formation in the inner layer**

According to the revised lone pair model,[28] the asymmetrically layered crystal structure is attributed to the stereo-chemically active lone pairs, where the filled s$^2$ electrons in the lower oxidation state of the cation show stereochemical activity that arises from the cation s–anion p coupling (anti-bonding states). The distorted crystal structure is dependent on the strength of the interaction between the cation s states and the anion p states, that is, on their relative energies. Therefore, the relatively low-symmetry crystal structures of γ-GeSe (*P6$_3$mc*) arise from the Ge 4s–Se 4p coupling, which is in contrast to the high-symmetry rocksalt structure (*Fm$\bar{3}$m*) of PbSe and PbS, where the energy of the Pb 6s electrons is so far away from the energies of Se 4p and S 3p electrons that the 6s$^2$ electrons are certainly stereo-chemically inert.

The lone-pair state of Ge in γ-GeSe can be visualized from the isosurface plots of the electron localization function (ELF), as shown in Fig. 3a-b. The projected density of states (PDOS) in Fig. 3c confirms the orbital hybridization of Ge 4s, Ge 4p, and Se 4p states in the blue-shaded regions, where the little Ge 4p shows a net stabilizing effect and is essential for Ge on-site hybridization while the Ge 4s lone-pairs contribute significantly at the anti-bonding level. In addition, there is also a strong Ge 4s state component in the bottom-most conduction band of γ-GeSe, which may be responsible for the localized electrons around Se atoms, as observed in the ELF plot (Fig. 3a). The strong Ge 4s–Se 4p antibonding coupling near the VBM of γ-GeSe is also reflected from the analysis of k-dependent crystal orbital Hamilton populations (COHPs) based on the LOBSTER code in PBE method.[29-32] In Fig. 3d, the positive/negative values in COHPs represent the antibonding/bonding state and the magnitude indicates the strength of orbital coupling.

Based on the above results, we summarized the schematic energy level diagram of the interactions in γ-GeSe in Fig. 3e. The occupied anti-bonding state consisting of Ge 4s and Se 4p coupling does not gain electronic energy, thus tending to break the Ge-Se bond and form a Ge vacancy. This mechanism is akin to the s-p antibonding coupling in $CH_3NH_3PbI_3$ and p-d antibonding coupling in $CuInSe_2$.[33, 34] The electronic structure of γ-GeSe resembles a defect-tolerant electronic structure known to lead to shallow defects.[35]

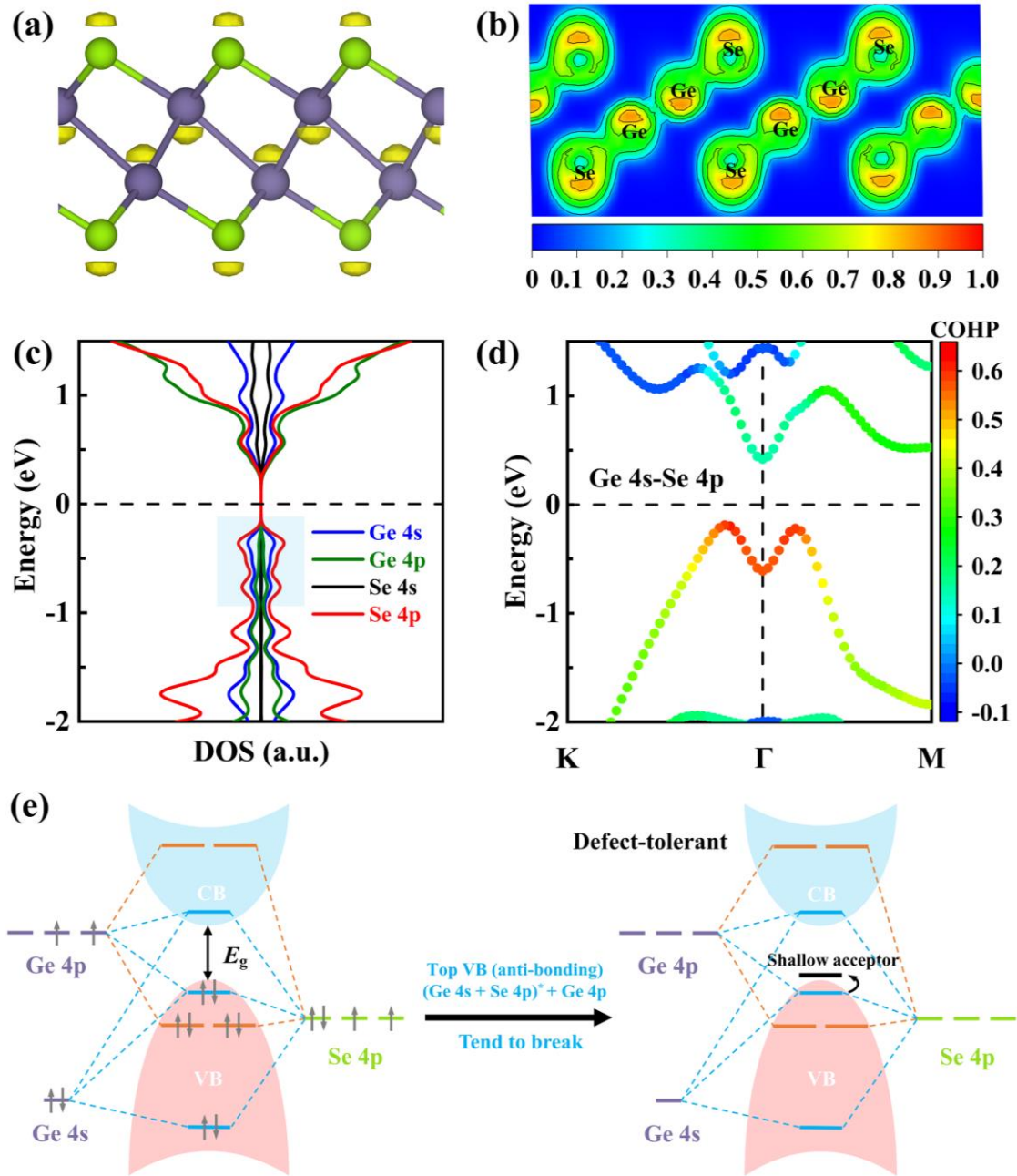

**Fig. 3** (a) ELF of γ-GeSe monolayer supercell (isosurface level: 0.813), showing the lone pairs in yellow. (b) 2D ELF contour map of the (110) plane in γ-GeSe. (c) PDOS of γ-GeSe using PBE functional. (d) The k-dependent COHPs of Ge 4s–Se 4p orbitals of γ-GeSe monolayer. (e) Schematic energy level diagram of the interactions in γ-GeSe and the tendency of forming a Ge vacancy and its associated shallow acceptor level. The Fermi level is set to 0.

Using first-principles calculations, Luo et al.[11] demonstrated that the gamma-phase of group-IV monochalcogenides (γ-MX, M = Ge, Sn; X = S, Te) can be stabilized in monolayer limit. Therefore, we conducted a calculation to explore whether the abnormal behavior of cationic vacancy also exists in these γ-MXs with the same symmetric structure. The $E_f$ of vacancies is calculated in the same method as γ-GeSe and the results are plotted as Fig. 4. It is found that the preferred formation of inner cationic vacancy is not unique to γ-GeSe, but also applicable to these γ-MXs. The k-dependent COHPs shows that there is M 4s–X 4p antibonding coupling near the VBM of all these γ-MXs. In addition, this abnormal behavior in these γ-MXs is positively related to the strength of anti-bonding states. The $E_f$ of cationic vacancy in GeS and SnS with strong anti-bonding state is significantly lower than that of GeTe and SnTe with weak anti-bonding state. And the energy difference ($E_{f\_mim}$) between the minimum $E_f$ of cationic and anionic vacancy decreases with the weakening of anti-bonding state. The above results indicate that the presence of the antibonding lone-pair state indeed accounts for the abnormal behavior of preferred formation of cationic vacancy from the interior, especially in γ-MX (M=Sn, Ge; X=S, Se, Te).

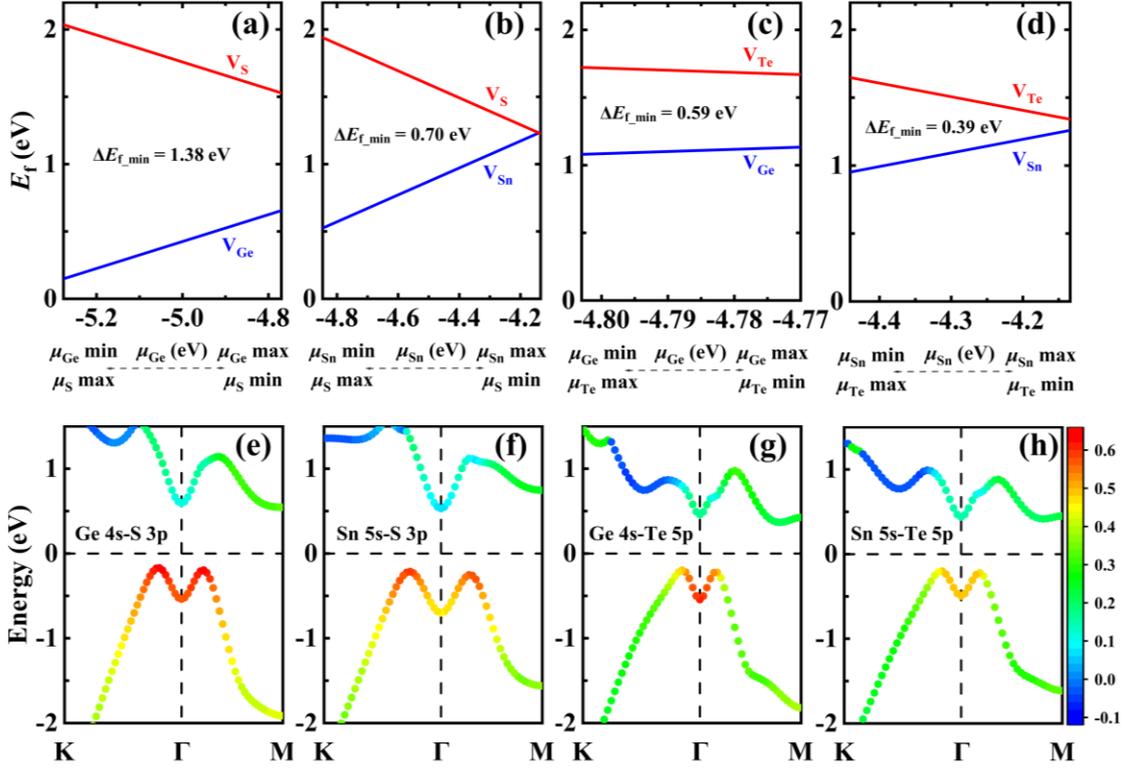

**Fig. 4** Formation energies of monolayer γ-GeS (a), γ-SnS (b), γ-GeTe (c), γ-SnTe (d) containing vacancies as functions of $\mu_{cation}$. The k-dependent COHPs of cation 4s–anion 4p orbitals of monolayer γ-GeS (e), γ-SnS (f), γ-GeTe (g), γ-SnTe (h) using PBE functional.

**Effects of $V_{Ge}$ on the electronic properties and light absorption.**

The diffusion energy barrier ($E_b$) and pathway to the neighboring site of $V_{Ge}$ are plotted in Fig. 5a. The $E_b$ of $V_{Ge}$ (1.04 eV) is much larger than that of the phosphorus vacancy in phosphorene (0.30 eV) but lower than that of the sulfur vacancy (2.27 eV) in MoS$_2$, indicating that it has moderate activity which can facilitate experimental observation.[36] To figure out the effect of $V_{Ge}$ on the electronic properties, we analyzed the ELF results, unfolded band structures, and local density of states (LDOS) for $V_{Ge}$-GeSe as well as for the perfect γ-GeSe. The ELF results in Fig. 5b show that the Ge and Se atoms around the $V_{Ge}$ give rise to distinct localized electrons relative to the perfect supercell (Fig. S6a-c), which may alter their interaction with gas molecules for gas sensing. As shown in Fig. S6d, the perfect γ-GeSe monolayer is a semiconductor with an indirect bandgap of 0.99 eV in HSE06 hybrid functional. However, the presence of $V_{Ge}$ in γ-GeSe (Fig. 5c) pushes upwardly the topmost valence band with its energy exceeding the Fermi level ($E_F$) and forming a shallow acceptor level, resulting in an insulator-to-metal transition and p-type nature. The LDOS of $V_{Ge}$-GeSe indicates that the defect state is composed of the contributions from both the Ge and Se atoms. The above results prove that the nature of the shallow acceptor level of $V_{Ge}$ is attributed to the special antibonding character (s-p coupling) at the top valence bands, which pushes the topmost valence band upwards to form a shallow acceptor level.

Since changes in the electronic structures can have a significant impact on the optical properties, the light absorption of γ-GeSe monolayers with and without $V_{Ge}$ are calculated, and the results are plotted in Fig. 5d. The perfect γ-GeSe monolayer exhibits strong absorption in the range from ultraviolet to visible light, while the presence of $V_{Ge}$ results in a significant change in the absorption spectrum, where it shows a strong peak in the infrared range due to the shallow intermediate gap state extending to the edge of the topmost valence band. In addition, due to the weak Coulomb screening effect in layered materials[37, 38] and considering the superb light utilization at the near-infrared region of its cousin phases[39], it is highly appealing to explore the excitonic behavior of this new hexagonal polymorph of GeSe in the future.

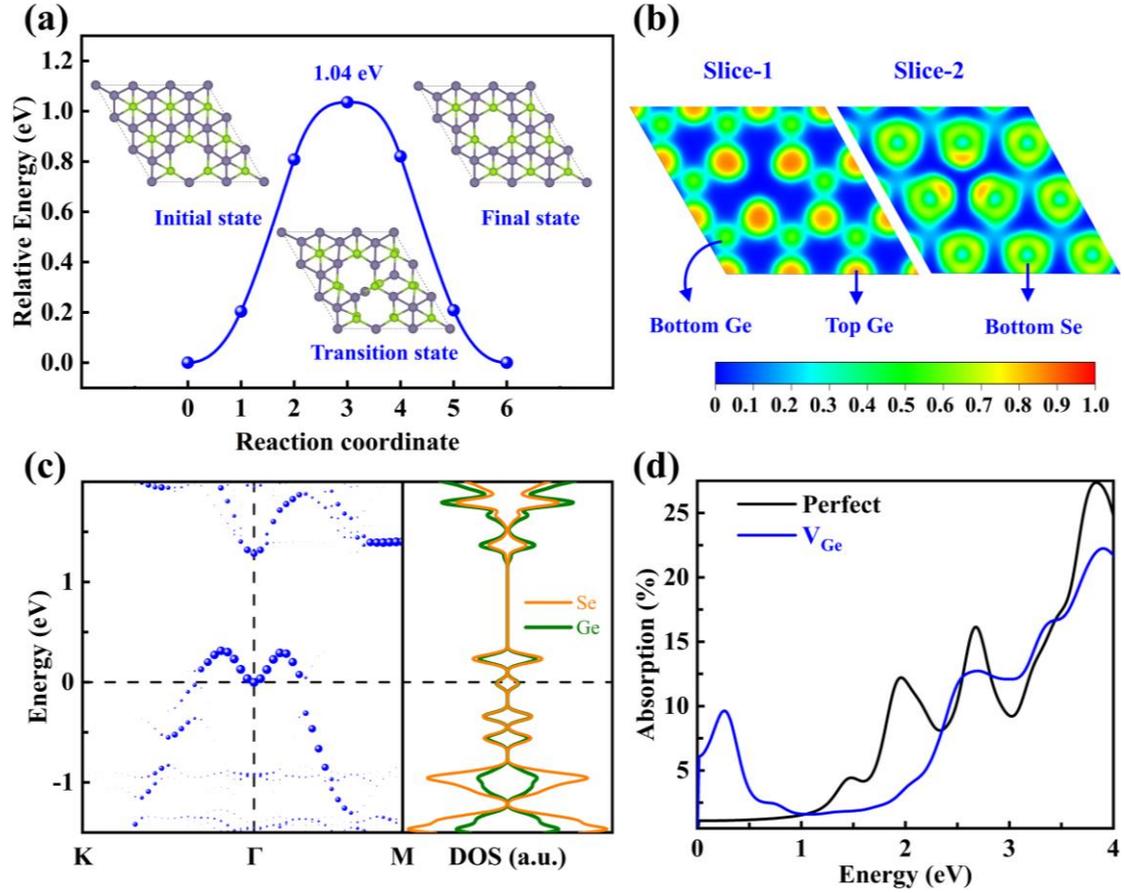

**Fig. 5** (a) Configurations and diffusion barrier of $V_{Ge}$, (b) 2D ELF contour maps of the (001) plane, (c) unfolded band structure and LDOS of monolayer $V_{Ge}$-GeSe, and (d) absorption spectra of monolayer γ-GeSe with and without $V_{Ge}$. The Fermi level is set to 0.

## Conclusions

In summary, we reveal the abnormal behavior of atomic defects in the recently synthesized new hexagonal phase of GeSe, as well as theoretical predicted hexagonal phase MX (M=Sn, Ge; X=S, Te). Different from the common fact that atomic defects, especially vacancies, are predominantly found at the surface due to the lower atomic coordination number, here, we demonstrate that γ-GeSe is a unique system showing the abnormal preference of forming Ge vacancies in the central region. This is in stark contrast to the formation of surface defects in common 2D compounds such as transition metal dichalcogenides, which are predominantly anionic vacancies. Our analysis of the orbital hybridization diagram suggests that the strong Ge 4s–Se 4p antibonding coupling near the VBM of γ-GeSe largely accounts for the tendency of breaking the Ge-Se bond, and accordingly, the relatively facile removal of Ge. The subsurface Ge vacancy located in the middle cationic Ge sublattice, accompanied by a shallow acceptor defect level, is well protected by the surface Se layers – a feature that is absent in other atomically thin materials. This renders the formation of sealed voids inside the layer which may be harnessed for quantum emitters and sensors, as well as for conducting filaments in neuromorphic applications with the agglomeration of Ge vacancy clusters.

## Methods

First-principles calculations were carried out by the Vienna ab-initio simulation package (VASP) with the projector augmented wave (PAW) method.[40, 41] The generalized gradient approximation (GGA) with the Perdew−Burke−Ernzerhof (PBE) functional was employed to describe the exchange-correlation energy.[42] The defective γ-GeSe monolayers were constructed based on a 3 × 3 × 1 supercell with a vacuum layer thickness larger than 20 Å. The energy cutoff and gamma centered k-meshes were set to be 400 eV and 5 × 5 × 1 for structural optimization, respectively. The convergence threshold for the residual force and total energy were 0.005 eV/Å and $10^{-6}$ eV, respectively. The DFT-D3 method was employed to describe the weak van der Waals interaction.[43, 44] The Heyd−Scuseria−Ernzerhof (HSE06) hybrid functional[45] and VASPKIT code[46] were used to evaluate the unfolded electronic structures of the supercell. Thermal and dynamic stability were verified by ab initio molecular dynamics (AIMD)[47] simulations at 300 K with a time step of 2 fs and phonon dispersion calculations based on the PHONOPY code,[48] respectively. To evaluate the diffusion energy barriers of defects, the climbing-image nudged elastic band (CI-NEB) method was performed.[49]

## Author contributions

Huan CM performed the calculations and wrote the original draft. Kripalani DR and Zhou K assisted in the analysis and drafted the manuscript. Cai YQ and Ke QQ guided the idea and finalized the manuscript. All authors participated in general discussion and the manuscript revision.

## Conflict of interest

The authors declare that they have no conflict of interest.

## Acknowledgements

This work was supported by the 100 Talents Program of Sun Yat-sen University (Grant 76220-18841201), the Natural Science Foundation of China (Grant 22022309) and Natural Science Foundation of Guangdong Province, China (2021A1515010024), the University of Macau (SRG2019-00179-IAPME, MYRG2020-00075-IAPME) and the Science and Technology Development Fund from Macau SAR (FDCT-0163/2019/A3).

## References

1. Y. Yu, X. Xu, Y. Wang, B. H. Jia, S. Huang, X. B. Qiang, B. Zhu, P. J. Lin, B. B. Jiang, S. X. Liu, X. Qi, K. F. Pan, D. Wu, H. Z. Lu, M. Bosman, S. J. Pennycook, L. Xie and J. Q. He, *Nat. Commun.*, 2022, **13**, 5612.

2. H. Jang, J. H. Park, H. S. Lee, B. Ryu, S. D. Park, H. A. Ju, S. H. Yang, Y. M. Kim, W. H. Nam, H. Wang, J. Male, G. J. Snyder, M. Kim, Y. S. Jung and M. W. Oh, *Adv. Sci.*, 2021, **8**, 2100895.

3. Z. Shu and Y. Q. Cai, *J. Mater. Chem. A*, 2021, **9**, 16056-16064.

4. S. Li, Z. C. Zhao, J. B. Li, H. Liu, M. S. Liu, Y. Q. Zhang, L. Z. Su, A. I. Perez-Jimenez, Y. C. Guo, F. Yang, Y. Liu, J. Z. Zhao, J. M. Zhang, L. D. Zhao and Y. H. Lin, *Small*, 2022, **18**, 2202507.

5. O. J. Clark, I. Wadgaonkar, F. Freyse, G. Springholz, M. Battiato and J. Sanchez-Barriga, *Adv. Mater.*, 2022, **34**, 2200323.

6. K. Jeong, H. Lee, C. Lee, L. H. Wook, H. Kim, E. Lee and M. H. Cho, *Appl. Mater. Today*, 2021, **24**, 101122.

7. S. C. Liu, C. M. Dai, Y. Min, Y. Hou, A. H. Proppe, Y. Zhou, C. Chen, S. Chen, J. Tang, D. J. Xue, E. H. Sargent and J. S. Hu, *Nat. Commun.*, 2021, **12**, 670.

8. B. Yan, B. Ning, G. X. Zhang, D. H. Zhou, X. Shi, C. X. Wang and H. Q. Zhao, *Adv. Opt. Mater.*, 2022, **10**, 2102413.

9. Z. Guan, Y. Zhao, X. Wang, N. Zhong, X. Deng, Y. Zheng, J. Wang, D. Xu, R. Ma, F. Yue, Y. Cheng, R. Huang, P. Xiang, Z. Wei, J. Chu and C. Duan, *ACS Nano*, 2022, **16**, 1308−1317.

10. S. Lee, J. E. Jung, H. G. Kim, Y. Lee, J. M. Park, J. Jang, S. Yoon, A. Ghosh, M. Kim, J. Kim, W. Na, J. Kim, H. J. Choi, H. Cheong and K. Kim, *Nano Lett.*, 2021, **21**, 4305-4313.

11. N. Luo, W. Duan, B. I. Yakobson and X. Zou, *Adv. Funct. Mater.*, 2020, **30**, 2000533.

12. C. M. Huan, P. Wang, B. H. He, Y. Q. Cai and Q. Q. Ke, *2D Mater.*, 2022, **9**, 045014

13. H.-g. Kim and H. J. Choi, *J. Mater. Chem. C*, 2021, **9**, 9683-9691.

14. C. M. Huan, P. Wang, B. T. Liu, B. H. He, Y. Q. Cai and Q. Q. Ke, *J. Mater. Chem. C*, 2022, **10**, 10995-11004.

15. Z. Shu, X. Y. Cui, B. W. Wang, H. J. Yan and Y. Q. Cai, *Chemsuschem*, 2022, **15**, 202200564.

16. C. M. Huan, P. Wang, B. H. He, Y. Q. Cai and Q. Q. Ke, *J. Mater. Chem. C*, 2022, **10**, 1839-1849.

17. W. Xu, L. Gan, R. Wang, X. Wu and H. Xu, *ACS Appl. Mater. Interfaces*, 2020, **12**, 19110-19115.

18. L. Liang, X. Li, Y. Sun, Y. Tan, X. Jiao, H. Ju, Z. Qi, J. Zhu and Y. Xie, *Joule*, 2018, **2**, 1004-1016.

19. R. Wang, Y. Su, G. Yang, J. Zhang and S. Zhang, *Chem. Mater.*, 2020, **32**, 1545-1552.

20. V. K. Sangwan, D. Jariwala, I. S. Kim, K. S. Chen, T. J. Marks, L. J. Lauhon and M. C. Hersam, *Nat. Nanotechnol.*, 2015, **10**, 403-406.

21. H. F. Chen, H. J. Yan and Y. Q. Cai, *Chem. Mater., 2022, **34**, 1020-1029*.

22. A. F. Kohan, G. Ceder, D. Morgan and C. G. V. d. Walle, *Phys. Rev. B*, 2000, **61**, 15019.

23. D. Edelberg, D. Rhodes, A. Kerelsky, B. Kim, J. Wang, A. Zangiabadi, C. Kim, A. Abhinandan, J. Ardelean, M. Scully, D. Scullion, L. Embon, R. Zu, E. J. G. Santos, L. Balicas, C. Marianetti, K. Barmak, X. Zhu, J. Hone and A. N. Pasupathy, *Nano Lett.*, 2019, **19**, 4371-4379.

24. M. B. Kanoun, *Surf. Interfaces*, 2021, **27**, 101442.

25. W. Zhou, X. Zou, S. Najmaei, Z. Liu, Y. Shi, J. Kong, J. Lou, P. M. Ajayan, B. I. Yakobson and J. C. Idrobo, *Nano Lett.*, 2013, **13**, 2615-2622.

26. D. D. Cuong, B. Lee, K. M. Choi, H. S. Ahn, S. Han and J. Lee, *Phys. Rev. Lett.*,

2007, **98**, 129903.

27. G. D. Lee, C. Z. Wang, E. Yoon, N. M. Hwang, D. Y. Kim and K. M. Ho, *Phys. Rev. Lett.*, 2005, **95**, 205501.

28. A. Walsh, D. J. Payne, R. G. Egdell and G. W. Watson, *Chem. Soc. Rev.*, 2011, **40**, 4455–4463.

29. V. L. Deringer, A. L. Tchougreeff and R. Dronskowski, *J. Phys. Chem. A*, 2011, **115**, 5461-5466.

30. R. Dronskowski and P. E. Bloechl, *J. Phys. Chem. C*, 1993, **97**, 8617–8624.

31. S. Maintz, V. L. Deringer, A. L. Tchougreeff and R. Dronskowski, *J. Comput. Chem.*, 2016, **37**, 1030-1035.

32. X. Sun, X. Li, J. Yang, J. Xi, R. Nelson, C. Ertural, R. Dronskowski, W. Liu, G. J. Snyder, D. J. Singh and W. Zhang, *J. Comput. Chem*, 2019, **40**, 1693-1700.

33. W. J. Yin, T. T. Shi and Y. F. Yan, *Appl. Phys. Lett.*, 2014, **104**, 063903.

34. C. Rincon and R. Marquez, *J. Phys. Chem. Solids*, 1999, **60**, 1865-1873.

35. R. E. Brandt, J. R. Poindexter, P. Gorai, R. C. Kurchin, R. L. Z. Hoye, L. Nienhaus, M. W. B. Wilson, J. A. Polizzotti, R. Sereika, R. Zaltauskas, L. C. Lee, J. L. MacManus-Driscoll, M. Bawendi, V. Stevanovic and T. Buonassisi, *Chem. Mater.*, 2017, **29**, 4667-4674.

36. Y. Cai, Q. Ke, G. Zhang, B. I. Yakobson and Y. W. Zhang, *J. Am. Chem. Soc.*, 2016, **138**, 10199-10206.

37. S. Refaely-Abramson, D. Y. Qiu, S. G. Louie and J. B. Neaton, *Phys. Rev. Lett.*,

2018, **121**, 167402.

38. C. Long, Y. Dai and H. Jin, *Phys. Rev. B*, 2021, **104**, 125306.

39. B. Mukherjee, Y. Cai, H. R. Tan, Y. P. Feng, E. S. Tok and C. H. Sow, *ACS Appl. Mater. Interfaces*, 2013, **5**, 9594-9604.

40. G. Kresse and J. Furthmuller, *Phys. Rev. B*, 1996, **54**, 11169-11186.

41. G. Kresse and J. Furthmuller, *Comput. Mater. Sci.*, 1996, **6**, 15-50.

42. J. P. Perdew, K. Burke and M. Ernzerhof, *Phys. Rev. Lett.*, 1996, **77**, 3865-3868.

43. S. Grimme, S. Ehrlich and L. Goerigk, *J. Comput. Chem.*, 2011, **32**, 1456-1465.

44. S. Grimme, J. Antony, S. Ehrlich and H. Krieg, *J. Chem. Phys.*, 2010, **132**, 154104.

45. J. Heyd, G. E. Scuseria and M. Ernzerhof, *J. Chem. Phys.*, 2003, **118**, 8207-8215.

46. V. Wang, N. Xu, J.-C. Liu, G. Tang and W.-T. Geng, *Comput. Phys. Commu.*, 2021, **267**, 108033.

47. G. J. Martyna, M. L. Klein and M. Tuckerman, *J. Chem. Phys.*, 1992, **97**, 2635-2643.

48. A. Togo and I. Tanaka, *Scripta. Mater.*, 2015, **108**, 1-5.

49. G. Henkelman, B. P. Uberuaga and H. Jonsson, *J. Chem. Phys.*, 2000, **113**, 9901-9904.